\documentclass[twocolumn, superscriptaddress]{revtex4-2}
\usepackage{graphicx}
\usepackage{xcolor}
\usepackage[setpagesize=false]{hyperref}
\hypersetup{colorlinks=true,linkcolor=[RGB]{0,0,128},citecolor=[RGB]{0,0,128},urlcolor=[RGB]{0,0,128}}

\usepackage{amsmath,amsbsy,amssymb}
\usepackage{bm}
\usepackage{mathrsfs}
\usepackage[normalem]{ulem}
\usepackage{textgreek}
\usepackage{booktabs}
\usepackage{physics2}
\usephysicsmodule{ab}
\usephysicsmodule{ab.braket}
\usephysicsmodule{diagmat}
\usepackage{fixdif}
\usepackage{derivative}

\newcommand{\Tr}{\operatorname{Tr}}

\newcommand{\mqty}[1]{\begin{matrix}#1\end{matrix}}

\newcommand{\Hc}{\mathrm{H.c.}}

\newcommand{\ua}{\uparrow}
\newcommand{\da}{\downarrow}
\newcommand{\dg}{\dagger}
\newcommand{\la}{\langle}
\newcommand{\ra}{\rangle}
\newcommand{\al}{\alpha}
\newcommand{\sg}{\sigma}
\newcommand{\gm}{\gamma}

\begin{document}

\title{
Halogen control of magnetic competition in Kitaev candidate Ru$X_3$ ($X =$ Cl, Br)
}

\author{Ryuta~Iwazaki}
\affiliation{
    Department of Physics, Graduate School of Science, Tohoku University, Sendai, Miyagi 980-8578, Japan
}

\author{Shinnosuke~Koyama}
\affiliation{
    Department of Physics, Tokyo Metropolitan University, Hachioji, Tokyo 192-0397, Japan
}

\author{Takashi~Koretsune}
\affiliation{
    Department of Physics, Graduate School of Science, Tohoku University, Sendai, Miyagi 980-8578, Japan
}

\author{Shintaro~Hoshino}
\affiliation{
    Department of Physics, Chiba University, Chiba 263-8522, Japan
}

\author{Joji~Nasu}
\affiliation{
    Department of Physics, Graduate School of Science, Tohoku University, Sendai, Miyagi 980-8578, Japan
}

\date{\today}

\begin{abstract}

    The spin-orbital Mott insulators Ru$X_3$ ($X =$ Cl, Br) have attracted considerable attention as promising candidate materials for realizing a Kitaev spin liquid.
    In this study, we construct effective pseudospin models from multiorbital Hubbard models derived from first-principles calculations and investigate the magnetic states of RuCl$_3$ and RuBr$_3$.
    From the constructed effective models, we find that RuBr$_3$ has more extended Wannier orbitals and stronger interlayer exchange interactions than RuCl$_3$.
    These interactions enhance three-dimensional correlations, consistent with the stronger antiferromagnetic tendency experimentally inferred for RuBr$_3$.
    Orbital-dependent Coulomb anisotropy further reduces the energy difference between ferromagnetic and zigzag states.
    Our results clarify how halogen substitution controls magnetic competition in Ru$X_3$ through interlayer exchange interactions and effects of orbital-dependent Coulomb interactions.

\end{abstract}

\maketitle

\section{Introduction}

Realizing a quantum spin liquid, a state of matter that does not exhibit magnetic ordering even at zero temperature, has been a long-standing challenge in condensed matter physics~\cite{Anderson1973, Fazekas1974}.
Thus far, candidate materials have been explored based on geometrically frustrated lattices, such as triangular and kagome lattices, where magnetic interactions are isotropic and the suppression of magnetic ordering arises from the lattice geometry~\cite{Ramirez1994}.
On the other hand, recent progress in materials design for quantum spin liquids has led to the proposal of a new route based on bond-dependent anisotropic interactions, which originate from orbital anisotropy and strong spin-orbit coupling (SOC)~\cite{Feiner1997,Khomskii2003}.
An ideal model for this route is the Kitaev model, which has an exact quantum spin liquid ground state~\cite{Kitaev2006}.
This model can be realized in materials where strong SOC generates an effective total angular momentum state $j_\mathrm{eff}=\frac 1 2$ at half filling and strong hybridization between the $d$ orbitals in magnetic ions and $p$ orbitals in ligand ions yields Ising-type interactions~\cite{Jackeli2009}.
Since the pioneering proposals for iridium compounds, the realization of Kitaev spin liquids has been suggested not only in $d$-electron systems, such as ruthenium- and cobalt-based compounds, but also in $f$-electron systems, including Yb compounds~\cite{Chaloupka2010, Singh2010, Yamaji2014prl, Rau2014, Winter2016, Liu2018, Sano2018, Jang2019, Stavropoulos2021}.
Interestingly, in the Kitaev quantum spin liquid, the elementary excitations are Majorana fermions, which are their own antiparticles and have been proposed as a platform for topological quantum computation.
Thus, extensive efforts have been devoted to materials design aimed at realizing a Kitaev spin liquid, and the search for such materials has become one of the central topics in condensed matter physics~\cite{Rau2016, Winter2017review, Matsuda2025}.

Among the various candidate materials for realizing a Kitaev spin liquid, $\al$-RuCl$_3$ (hereafter simply referred to as RuCl$_3$) has attracted significant attention as one of the most promising candidates~\cite{Plumb2014}.
This material, however, undergoes an antiferromagnetic transition at $T_\mathrm{N}\simeq 7\,\mathrm K$ due to the presence of a trigonal crystalline electric field (CEF) and direct $d$-orbital hybridization, which lead to an effective spin model that includes not only the Kitaev interaction but also Heisenberg and additional anisotropic interactions~\cite{Rau2014, Winter2016}.
Nevertheless, since the energy scale of the Kitaev interaction is believed to be dominant compared to the other interactions, characteristic features intrinsic to the Kitaev model have been experimentally observed above the N\'{e}el temperature: Raman and neutron-scattering experiments have revealed the presence of a broad excitation continuum, which has been proposed as a signature of fractionalized excitations in a proximate Kitaev spin liquid~\cite{Sandilands2015,Banerjee2017}.
Furthermore, the observation of half-integer quantization in the thermal Hall coefficient under magnetic fields~\cite{Kasahara2018prl,Kasahara2018} has been regarded as compelling evidence for a Majorana chiral edge mode, which is a hallmark of the Kitaev spin liquid under magnetic fields.
Recently, to extract the intrinsic properties of RuCl$_3$ as a Kitaev material, high-quality single crystals have been synthesized~\cite{Namba2024}, and it has been clarified that, at low temperatures, the Ru sites exhibit threefold rotational symmetry in the $R\bar 3$ structure~\cite{Park2024}, which is consistent with the idealized structure for realizing the Kitaev interaction.

To realize a quantum spin liquid, it is crucial to suppress magnetic ordering, and thus, understanding the origin of magnetic ordering is important for assessing how far a material deviates from the ideal Kitaev model. 
In RuCl$_3$, although the zigzag (ZZ) magnetic order is stabilized at low temperatures, an energetic competition between ferromagnetic (FM) and ZZ orders has been pointed out by resonant inelastic x-ray scattering (RIXS) experiments~\cite{Suzuki2021}, which suggests the presence of predominant FM Kitaev interactions.
Furthermore, in-plane magnetic fields have been found to induce an additional phase in the field region adjacent to the quantum spin-liquid phase.
The induced phase is interpreted as a ZZ order with a six-layer periodicity along the out-of-plane direction, which differs from the three-layer periodic magnetic structure along the out-of-plane direction in zero field~\cite{Balz2021}.
The emergence of this field-induced long-period ZZ order along the out-of-plane direction has highlighted the importance of interlayer interactions between the honeycomb planes~\cite{Janssen2020, Balz2021, Cen2025}.

Since the Kitaev interaction arises from hybridization with ligand orbitals, halogen substitution in RuCl$_3$ is expected to significantly affect the magnetic properties by modifying the orbital hybridization and the charge-transfer gap~\cite{Ersan2019}.
Based on this consideration, the synthesis of Ru$X_3$ ($X =$ Cl, Br, I) materials with halogen substitution has been successfully achieved, and their magnetic properties have been actively investigated~\cite{Nawa2021,Imai2022,Ni2022,Gretarsson2024,Pearce2024,Nawa2026,Gretarsson2026}, along with theoretical studies of these materials~\cite{Kim2021,Kaib2022}.
It has been reported that RuI$_3$ exhibits metallic behavior, whereas RuCl$_3$ and RuBr$_3$ are insulators; therefore, the latter two materials are more suitable for investigating magnetic properties as candidates for Kitaev materials~\cite{Banerjee2017, Imai2022, Nawa2021, Ni2022}.
Although RuBr$_3$ has been synthesized in the $R\bar{3}$ structure, which is the same crystal structure as RuCl$_3$, it undergoes an antiferromagnetic transition to a ZZ magnetic order at $T_\mathrm{N} \simeq 34\,\mathrm{K}$, which is higher than that of RuCl$_3$~\cite{Imai2022}.
This higher N\'{e}el temperature is consistent with the fact that the powder-averaged Curie-Weiss temperature obtained from magnetic susceptibility measurements in RuBr$_3$ is smaller than that in RuCl$_3$~\cite{Suzuki2021,Li2021,Imai2022}, suggesting that antiferromagnetic correlations are stronger in RuBr$_3$ than in RuCl$_3$~\cite{Kaib2022}.
However, the microscopic origin of the stronger antiferromagnetic correlations in RuBr$_3$ remains unclear.

To understand the magnetic properties of Ru$X_3$, previous theoretical studies have widely employed multiorbital Hubbard models as a starting point and have analyzed effective pseudospin models incorporating SOC~\cite{Rau2016,Winter2017review}.
In these studies, calculations have mainly been performed under the assumption of an ideal localized $j_\mathrm{eff}=\frac 1 2$ states derived from the $t_{2g}$ orbitals under a cubic crystal field in the infinite SOC limit.
However, in reality, the trigonal CEF and $dp$ hybridization, as well as the influence of the $j_\mathrm{eff}=\frac 3 2$ orbitals due to finite SOC, are expected to lead to deviations from the ideal $j_\mathrm{eff}=\frac 1 2$ state.
In particular, the replacement of Cl with Br is expected to reduce the charge-transfer gap and enhance $dp$ hybridization, which may lead to an extension of Wannier orbitals and thereby increase interlayer interactions~\cite{Imai2022,Gretarsson2024}.
Indeed, the recent experimental study using RIXS has pointed out that the interlayer interactions are enhanced by halogen substitution with Br, which motivates further investigations that take into account the spatial extent of the localized states~\cite{Gretarsson2026}.
Moreover, in the perturbation process for deriving the Kitaev interaction from the strong-coupling limit, Coulomb interactions within the $t_{2g}$ orbitals have mainly been considered as intermediate states~\cite{Rau2014,Winter2016}, while the influence of electron correlations derived from the $e_g$ orbitals has not been sufficiently taken into account.
This effect may also be important for understanding the differences in magnetic properties between RuCl$_3$ and RuBr$_3$, because the $e_g$ orbitals are expected to be more involved in the low-energy physics of RuBr$_3$ than in RuCl$_3$, owing to the larger ratio of the Hund coupling to the on-site Coulomb interaction ($J/U$) in RuBr$_3$~\cite{Kaib2022}.

In this paper, to clarify the effects of halogen substitution on the magnetic states of Ru$X_3$ with $X = \mathrm{Cl}$ and $\mathrm{Br}$, we perform first-principles calculations and construct low-energy effective models based on perturbation theory in the strong-correlation limit, where the localized states are also determined from first-principles calculations.
We find that the orbital-averaged spread of the Wannier orbitals in RuBr$_3$ is larger than that in RuCl$_3$, leading to stronger interlayer interactions in RuBr$_3$ and enhanced three-dimensional antiferromagnetic correlations.
We then analyze the low-energy effective models within mean-field theory and show that the stable magnetic structure in RuCl$_3$ is the ZZ order, whereas the FM order is stabilized in RuBr$_3$.
This difference mainly arises from the suppression of the FM Kitaev interaction in RuBr$_3$ relative to that in RuCl$_3$.
Since experimental observations indicate that the magnetic structure in RuBr$_3$ is also the ZZ order, we further analyze a model incorporating anisotropic Coulomb interactions and find that the energy difference between the FM and ZZ orders is significantly reduced as the anisotropy increases.
These results show that halogen substitution controls magnetic competition in Ru$X_3$ through both interlayer exchange interactions and effects of orbital-dependent Coulomb interactions.

\section{Results}
\subsection{DFT calculation}
    \begin{figure}[t]
        \centering
        \includegraphics[width=0.66\linewidth]{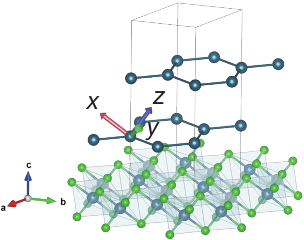}
        \caption{
            \textbf{Crystal structure of Ru$X_3$ with conventional hexagonal lattice illustrated by \texttt{VESTA}}~\cite{VESTA}.
            Blue spheres represent Ru atoms, and green spheres represent halogen atoms.
            For clarity, halogen sites in the second and third layers are omitted.
            We also show a local Cartesian coordinate $x$, $y$, and $z$ for a Ru$X_6$ octahedron.
        }
        \label{fig:crystal_structure}
    \end{figure}
    
    \begin{figure*}[t]
        \centering
        \includegraphics[width=\linewidth]{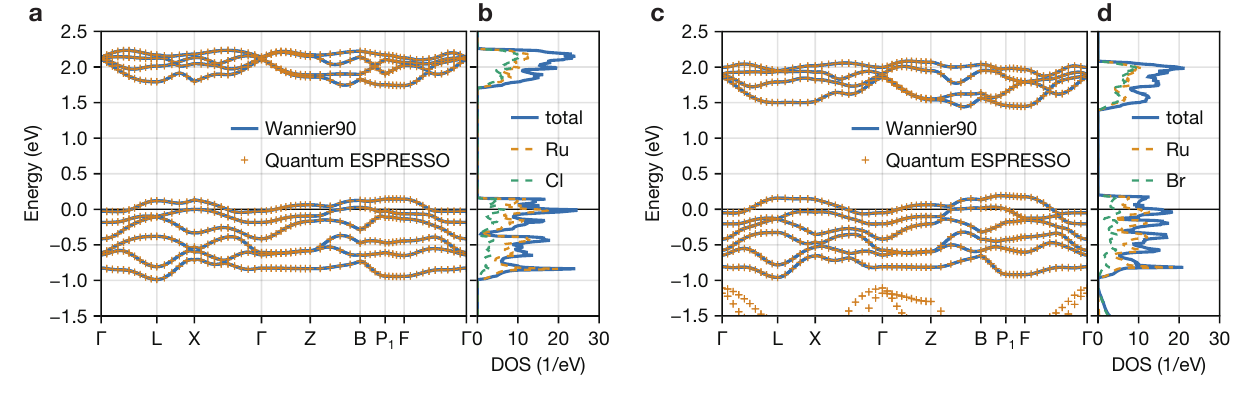}
        \caption{
            \textbf{Electronic structures of RuCl$_3$ and RuBr$_3$}.
            Energy band structures and densities of states computed by \texttt{Quantum ESPRESSO} and \texttt{Wannier90} for (a,b) RuCl$_3$ and (c,d) RuBr$_3$.
            The Fermi energy is set to 0 eV.
            In panels (a) and (c), yellow crosses denote the DFT results obtained by using \texttt{Quantum ESPRESSO}, and blue solid lines represent the bands fitted using Wannier functions from \texttt{Wannier90}.
            High-symmetry points along the $\bm k$-path follow Ref.~\cite{Setyawan2010}.
            In panels (b) and (d), total DOS and projected ones obtained from DFT are presented by solid and dashed lines, respectively.
        }
        \label{fig:pdos_band}
    \end{figure*}

In this section, we present the results of density functional theory (DFT) calculations and the construction of Wannier functions based on these calculations for RuCl$_3$ and RuBr$_3$ with the $R\bar 3$ structure (see Fig.~\ref{fig:crystal_structure}), as reported in Refs.~\cite{Park2024, Imai2022}.
Figure~\ref{fig:pdos_band} shows the band structures and densities of states (DOS) for these materials obtained from generalized gradient approximation (GGA)+SOC calculations using \texttt{Quantum ESPRESSO}.
Figures~\ref{fig:pdos_band}(a) and (c) present the band structures of RuCl$_3$ and RuBr$_3$, respectively.
In these panels, we show the bands fitted using Wannier functions constructed with \texttt{Wannier90}, using atomic $d$-orbital wave functions centered on the Ru sites as initial projections (see Method).
The excellent agreement between the DFT results and the Wannier-fitted bands suggests that the Wannier functions are properly constructed using only the Ru $d$ orbitals.
We also performed GGA+SOC calculations and Wannierization for RuI$_3$ in the $R\bar 3$ structure~\cite{Ni2022}.
However, we could not construct Wannier functions using only Ru $d$ orbitals, in contrast to the cases of RuCl$_3$ and RuBr$_3$, possibly due to the strong hybridization between the I $5p$ orbitals and the Ru $4d$ orbitals.
We consider that an analysis explicitly including the iodine $5p$ orbitals is required to construct Wannier functions for RuI$_3$, as has been performed for CrI$_3$~\cite{Stavropoulos2021}.

Here, we discuss the differences in the band structures between RuCl$_3$ and RuBr$_3$.
Figures~\ref{fig:pdos_band}(b) and (d) show the density of states (DOS) for RuCl$_3$ and RuBr$_3$, respectively.
In both materials, the bands in the energy range $-1.0 \lesssim E \lesssim 2.5$ around the Fermi level mainly originate from Ru $d$ orbitals.
From the orbital-resolved DOS calculated using the constructed Wannier functions, we find that the bands in the range $-1.0 \lesssim E \lesssim 0.3$ near the Fermi level are dominated by $t_{2g}$ orbitals, while those in the range $1.5 \lesssim E \lesssim 2.5$ are mainly attributed to $e_g$ orbitals.
From these figures, the energy splitting between the two sets of bands in RuBr$_3$, which corresponds to the crystal-field splitting between the $t_{2g}$ and $e_g$ orbitals, is smaller than that in RuCl$_3$.
This reduction of the $t_{2g}$--$e_g$ splitting is consistent with previous GGA+SOC calculations for Ru$X_3$~\cite{Nawa2021}.
The absolute energy scale is also compatible with spectroscopic estimates: photoemission studies suggest a $t_{2g}$--$e_g$ splitting of approximately $2.2~\mathrm{eV}$ in RuCl$_3$~\cite{Sinn2016,Sandilands2016,Koitzsch2016}, while RIXS measurements find the corresponding excitation in RuBr$_3$ around $2.0~\mathrm{eV}$~\cite{Gretarsson2024}.
In addition, as shown in Fig.~\ref{fig:pdos_band}(d), bands mainly originating from Br $4p$ orbitals are located just below the bands dominated by Ru $d$ orbitals in the energy region below $-1.0$\,eV, indicating a smaller charge-transfer gap in RuBr$_3$ than in RuCl$_3$.
Indeed, the orbital-averaged spreads of the constructed Wannier functions are $3.58\,\text{\AA}^2$ for RuCl$_3$ and $4.78\,\text{\AA}^2$ for RuBr$_3$.
This larger spread in RuBr$_3$ possibly originates from stronger hybridization between Ru $4d$ orbitals and ligand $p$ orbitals, which is attributed to the smaller charge-transfer gap in RuBr$_3$ compared to RuCl$_3$.

\subsection{Multiorbital Hubbard model}

\begin{figure*}[t]
    \centering
    \includegraphics[width=\linewidth]{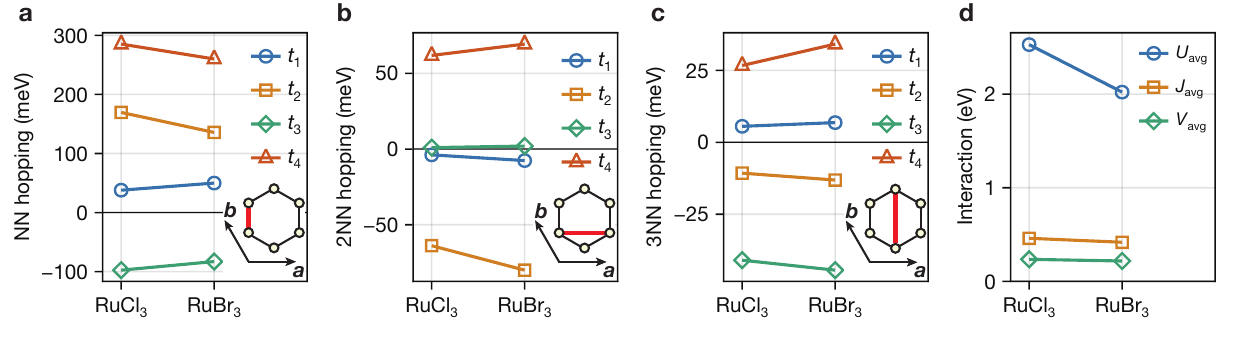}
    \caption{
        \textbf{Calculated transfer integrals and Coulomb interaction parameters for RuCl$_3$ and RuBr$_3$}.
        The in-plane transfer integrals in RuCl$_3$ and RuBr$_3$ are shown for (a) nearest-neighbor (NN), (b) second-nearest-neighbor (2NN), and (c) third-nearest-neighbor (3NN) bonds within a honeycomb plane.
        Blue circles, yellow squares, green diamonds, and red triangles correspond to $t_1 = t_{yz,yz}, t_2 = t_{yz, zx}, t_3 = t_{xy, xy}, t_4 = t_{z^2, xy}$, respectively, for the Z-bond indicated by the solid red line in the inset of each panel.
        The plotted hopping parameters are spin-independent quantities obtained by averaging the spin-conserving matrix elements, $t_{\gm\gm'} = (t_{\gm\gm'}^{\ua\ua} + t_{\gm\gm'}^{\da\da})/2$.
        The local $x, y, z$ coordinates are shown in Fig.~\ref{fig:crystal_structure}.
        These hopping parameters correspond to the dominant transfer integrals commonly used to characterize Kitaev materials.
        (d) Orbital-averaged Coulomb interaction parameters.
        Blue circles, yellow squares, and green diamonds represent $U_\mathrm{avg}, J_\mathrm{avg}, V_\mathrm{avg}$, respectively.
        See Eqs.~\eqref{eq:def_Uavg}--\eqref{eq:def_Vavg} for the definitions.
    }
    \label{fig:hopping}
\end{figure*}

Based on the Wannier functions constructed in the previous section, we derive a multiorbital Hubbard model for Ru$X_3$.
Figures~\ref{fig:hopping}(a)--\ref{fig:hopping}(c) show the parameters of transfer integrals for in-plane nearest-neighbor (NN), second-nearest-neighbor (2NN), and third-nearest-neighbor (3NN) bonds, respectively, obtained using \texttt{Wannier90}.
We focus on the four representative hopping integrals commonly discussed in Kitaev materials, $t_1=t_{yz,yz}$, $t_2=t_{yz,zx}$, $t_3=t_{xy,xy}$, and $t_4=t_{z^2,xy}$, for the Z-bond, which is one of the three types of NN bonds perpendicular to the $z$ axis (see Fig.~\ref{fig:crystal_structure}).
The overall behavior of the transfer integrals is consistent with results reported in a previous study using \texttt{VASP}~\cite{Kaib2022}.
In particular, the value of $t_2$ decreases from RuCl$_3$ to RuBr$_3$, which may lead to the suppression of the FM Kitaev interaction in RuBr$_3$ relative to that in RuCl$_3$.
Although this behavior is not intuitive because $t_2$ is ascribed to the mixing with ligand orbitals, which is expected to be stronger in RuBr$_3$ as discussed above, note that a similar trend has also been reported in a previous study using a different DFT code~\cite{Kim2021,Kaib2022}.
We also find that the amplitudes of the transfer integrals for the 2NN and 3NN bonds in RuBr$_3$ are slightly larger than those in RuCl$_3$.
This can be understood from the increased spatial extent of the Wannier functions, as mentioned previously.
Moreover, $t_4$ for the NN bond is larger than the other hopping parameters, including $t_2$, which results in the Kitaev interaction.
This transfer integral $t_4$ represents hybridization between $t_{2g}$ and $e_g$ orbitals, and its large magnitude suggests a significant contribution of $e_g$ orbitals to the low-energy physics of Ru$X_3$.

Figure~\ref{fig:hopping}(d) shows the Coulomb interaction parameters calculated using the constrained random phase approximation (cRPA) with \texttt{RESPACK}.
We compute the direct and exchange integrals for the Wannier orbitals and define the orbital-averaged parameters as follows.
\begin{align} \label{eq:def_Uavg}
    &U_\mathrm{avg}
    = \frac{1}{(2\ell + 1)^2} \sum_{\gm\gm'} U_{\gm\gm'},
    \\ \label{eq:def_Javg}
    &U_\mathrm{avg} - J_\mathrm{avg}
    = \frac{1}{2\ell (2\ell + 1)} \sum_{\gm\neq\gm'} \ab( U_{\gm\gm'} - J_{\gm\gm'} ),
    \\ \label{eq:def_Vavg}
    &V_\mathrm{avg}
    = \frac{1}{\sum_{\gm\gm', \bm \delta \in \mathrm{NN}} 1} \sum_{\gm\gm', \bm \delta \in \mathrm{NN}} U_{\gm\gm';\bm{\delta}},
\end{align}
where $\ell = 2$, and $U_{\gm\gm'}$ and $J_{\gm\gm'}$ denote the on-site direct and exchange integrals between orbitals $\gm$ and $\gm'$ at the same site, respectively, while $U_{\gm\gm';\bm{\delta}}$ represents the direct integral between orbital $\gm$ at one site and orbital $\gm'$ at a neighboring site separated by $\bm{\delta}$.
As shown in Fig.~\ref{fig:hopping}(d), the onsite interaction $U_\mathrm{avg}$ in RuBr$_3$ is smaller than that in RuCl$_3$, which is consistent with the larger spatial extent of the Wannier functions in RuBr$_3$.
In contrast, the Hund coupling $J_\mathrm{avg}$ does not differ significantly between the two materials, suggesting that the ratio $J_\mathrm{avg}/U_\mathrm{avg}$ is larger in RuBr$_3$ than in RuCl$_3$.
The parameter $V_\mathrm{avg}$ represents the strength of the Coulomb interaction between NN sites, and its value is much smaller than $U_\mathrm{avg}$, supporting the use of a multiorbital Hubbard model that includes only local Coulomb interactions as a first approximation.
We note that Coulomb interactions are generally expressed as four-center integrals $U_{\gm_1\gm_2\gm_3\gm_4}$.
In the present study, we construct the multiorbital Hubbard model using spherically symmetrized Coulomb interactions derived from four-center Coulomb integrals computed by \texttt{RESPACK} (see Methods).

\subsection{Pseudospin model} \label{sec:spin_model}
\begin{figure*}[t]
\centering
\includegraphics[width=\linewidth]{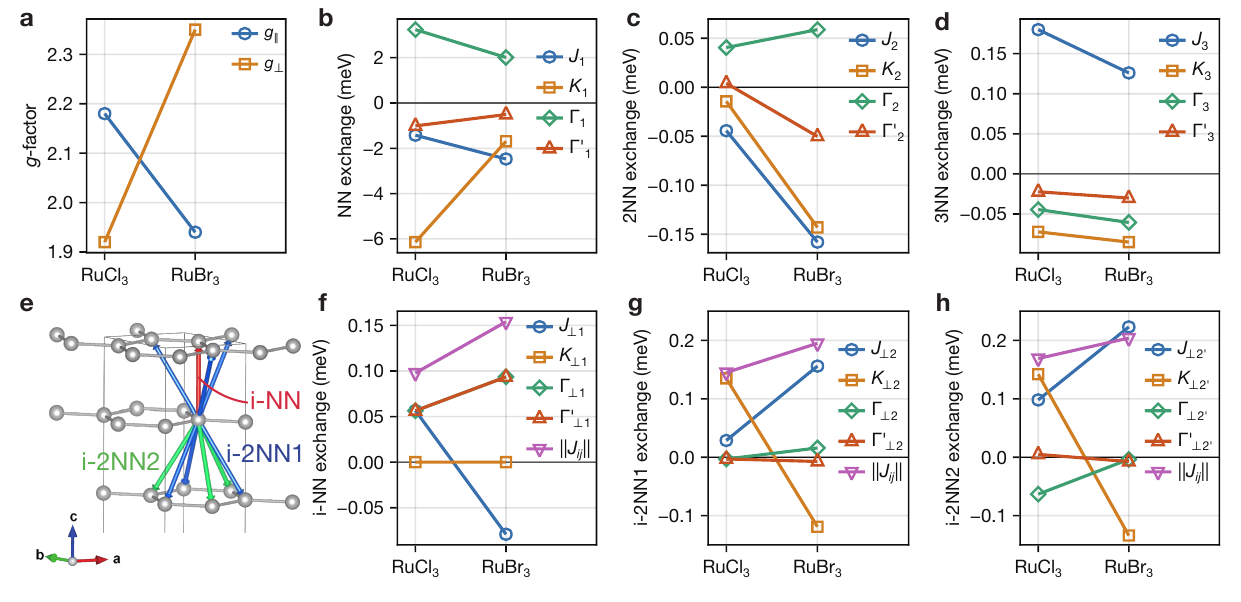}
\caption{
\textbf{Calculated parameters of the effective pseudospin model for RuCl$_3$ and RuBr$_3$}.
(a) Halogen dependence of the $g$-factors, where blue circles represent the in-plane components $g_\parallel$ and yellow squares represent the out-of-plane components $g_\perp$.
(b)-(d) Coupling constants for in-plane NN, 2NN, and 3NN Z-bonds, respectively.
The corresponding bond is indicated in the inset of Figs.~\ref{fig:hopping}(a)--(c).
Blue circles denote the Heisenberg interaction $J$, yellow squares denote the Kitaev interaction $K$, and green diamonds and red triangles denote the two types of anisotropic interactions $\Gamma$ and $\Gamma'$, respectively.
(e) Schematic illustration of the interlayer interactions, drawn using \texttt{VESTA}~\cite{VESTA}.
Red, blue, and green arrows correspond to interlayer NN (i-NN) bonds, interlayer second-NN bonds within the same sublattice (i-2NN1), and interlayer second-NN bonds between different sublattices (i-2NN2), respectively.
For clarity, an in-plane honeycomb network of Ru atoms is shown in this figure.
(f)-(h) Coupling constants for i-NN, i-2NN1, and i-2NN2, respectively.
Pink inverted triangles represent the matrix norm $||J_{ij}||$ of the coupling matrix.
}
\label{fig:coupling_constant}
\end{figure*}

From the multiorbital Hubbard model with parameters obtained from DFT calculations, we derive a low-energy effective model by performing second-order perturbation theory with respect to intersite hopping~\cite{Iwazaki2021, Iwazaki2023}.
This effective model is expressed in terms of pseudospin-$\frac 1 2$ operators $\tilde{\bm S}_i$, defined within the lowest Kramers doublet of the $d^5$ configuration, which is obtained by diagonalizing the local part of the multiorbital Hubbard model.
We note that the local basis for the pseudospin operators is not simply given by the $j_\mathrm{eff}=\frac 1 2$ states derived from the $t_{2g}$ orbitals under a cubic crystal field in the infinite SOC limit, but is instead determined by considering local contributions obtained from DFT calculations, such as deviations from the cubic crystal field and effects of finite SOC.
The effective Hamiltonian is formally given by $\mathscr H_\mathrm{eff} = \frac{1}{2}\sum_{ij}\sum_{\mu\nu} J_{ij}^{\mu\nu} \tilde S_i^\mu \tilde S_j^\nu$, where $J_{ij}^{\mu\nu}$ is a $3\times 3$ matrix of coupling constants on the bond connecting sites $i$ and $j$.
We also evaluate the Zeeman term, in which the pseudospin $\tilde{\bm S}_i$ is coupled to magnetic fields as $-\sum_i \sum_{\mu\nu} h^\mu g^{\mu\nu} \tilde S_i^\nu$, where $g$ is the $3\times 3$ $g$-factor tensor.
Figure~\ref{fig:coupling_constant}(a) shows the calculated $g$-factors for RuCl$_3$ and RuBr$_3$, where $g_\parallel$ and $g_\perp$ denote the components parallel and perpendicular to the honeycomb plane, respectively (see Methods for detailed derivations).
Our results indicate that RuCl$_3$ exhibits easy-plane anisotropy ($g_\parallel > g_\perp$), whereas RuBr$_3$ shows easy-axis anisotropy.
In experiments, both RuCl$_3$ and RuBr$_3$ exhibit easy-plane anisotropy, which is not reproduced in our calculations for RuBr$_3$~\cite{Kubota2015, Johnson2015, Pearce2024}.
The origin of this discrepancy for RuBr$_3$ will be discussed in Sec.~\ref{sec:aniso}.

Figures~\ref{fig:coupling_constant}(b)--\ref{fig:coupling_constant}(d) show the in-plane coupling constants obtained from second-order perturbation calculations, corresponding to NN, 2NN, and 3NN bonds, respectively.
To simplify the discussion, we introduce a conventional notation for the coupling constants.
For example, the exchange constant matrix for the Z-bond is written as
\begin{align} \label{eq:Jmel}
    J_{ij}^Z = \ab(\mqty{
        J & \Gamma & \Gamma' \\
        \Gamma & J & \Gamma' \\
        \Gamma' & \Gamma' & J + K
    }),
\end{align}
where $J$, $K$, $\Gamma$, and $\Gamma'$ represent the Heisenberg, Kitaev, and two types of off-diagonal interactions, respectively.
The detailed derivations are given in the Methods section.
The extended Kitaev model including these interactions within NN bonds has been widely studied as a minimal model for describing the magnetic properties of RuCl$_3$.
Here, we compare our results for the in-plane NN coupling constants with those of previous studies on the NN $JK\Gamma$ model~\cite{Rau2014}.
Their classical analyses indicate that the exchange constants for RuCl$_3$ obtained in our calculations stabilize an FM phase, whereas previous exact-diagonalization studies suggest a ZZ phase in the presence of quantum fluctuations.
Thus, our results for the in-plane NN coupling constants are consistent with experimental suggestions that quantum fluctuations play an important role in stabilizing the ZZ order in RuCl$_3$~\cite{Suzuki2021}.
On the other hand, based on previous theoretical studies of the NN $JK\Gamma$ model, the exchange constants for RuBr$_3$ lead to an FM phase even in the presence of quantum fluctuations, which is inconsistent with the experimental observation of a ZZ order in RuBr$_3$~\cite{Imai2022}.
Moreover, our results for longer range interactions beyond NN bonds indicate that absolute values of the coupling constants for second NN interactions are larger in RuBr$_3$ than in RuCl$_3$, particularly for the ferromagnetic Heisenberg and Kitaev interactions [Fig.~\ref{fig:coupling_constant}(c)], and the antiferromagnetic Heisenberg interaction for the third NN bond is suppressed in RuBr$_3$ [Fig.~\ref{fig:coupling_constant}(d)].
These results suggest that the longer-range interactions beyond NN bonds also stabilize the FM order.

As discussed above, it is difficult to explain the experimentally observed ZZ order in RuBr$_3$ solely on the basis of the in-plane coupling constants, implying the importance of interlayer interactions.
Figures~\ref{fig:coupling_constant}(f-h) show the coupling constants for the interlayer NN (i-NN) bond and two types of interlayer 2NN bonds, namely, those within the same sublattice of the honeycomb lattice (i-2NN1) and those between different sublattices (i-2NN2), as illustrated in Fig.~\ref{fig:coupling_constant}(e).
Note that these interactions on the three types of bonds are the dominant interlayer interactions, whereas the other interlayer interactions are significantly smaller.
Figure~\ref{fig:coupling_constant}(f) shows the exchange interactions on the i-NN bond.
The exchange couplings satisfy $K=0$ and $\Gamma = \Gamma'$ due to inversion symmetry~\cite{Cen2025}.
In this bond, the sign of the Heisenberg interaction becomes ferromagnetic in RuBr$_3$, while the other interactions remain antiferromagnetic and largely enhanced by changing the halogen atom from Cl to Br.
As an overall tendency, the magnitudes of the dominant interlayer interactions in RuBr$_3$ are larger than those in RuCl$_3$, which is consistent with the larger spatial extent of the Wannier functions in RuBr$_3$.
Since this enhancement of interlayer interactions may weaken the FM tendency, it could contribute to the stabilization of the ZZ order in RuBr$_3$, which will be discussed in more detail in the next paragraph.
To quantitatively examine the overall strength of the interlayer interactions, we define the matrix norm of the coupling matrix $J_{ij}$ as $||J_{ij}|| = \sqrt{\Tr[J_{ij}^\mathrm{T} J_{ij}]/3}$.
As presented in Fig.~\ref{fig:coupling_constant}(f-h), the norms of the coupling matrices in RuBr$_3$ are larger by a factor of approximately 1.2--1.5 for the three types of the interlayer bonds, suggesting that interlayer interactions play an important role in RuBr$_3$.
This result is also consistent with a recent experimental study using RIXS, which pointed out that interlayer spin correlations are enhanced by halogen substitution with Br~\cite{Gretarsson2026}.
The enhanced interlayer exchange therefore represents one of the key channels through which halogen substitution affects the magnetic competition in Ru$X_3$.

\begin{table}[t]
\centering
\caption{
In-plane and out-of-plane components of the Curie-Weiss temperature, $\Theta^\parallel$ and $\Theta^\perp$, and their average, $\Theta^\mathrm{ave}$.
``2D'' indicates calculations using only in-plane NN, 2NN, and 3NN interactions, whereas ``3D'' indicates calculations that additionally include the three interlayer interactions, i-NN, i-2NN1, and i-2NN2.
}
\begin{tabular}{ll|ccc} \toprule\toprule
$(\mathrm K)$ & & $\Theta^\parallel$ & $\Theta^\perp$ & $\Theta^\mathrm{ave}$
\\ \hline
RuCl$_3$ & 2D & 33.24 & 22.51 & 29.66 \\
        & 3D & 30.33 & 20.21 & 26.95
\\
RuBr$_3$ & 2D & 31.45 & 24.53 & 29.14 \\
        & 3D & 28.35 & 20.75 & 25.81
    \\ \bottomrule\bottomrule
\end{tabular}
\label{tab:CW_temp}
\end{table}

To further examine the effect of interlayer interactions on magnetic correlations, we evaluate the Curie-Weiss temperature using the modified Curie-Weiss law~\cite{Li2021}.
The Curie-Weiss temperatures of the in-plane and out-of-plane components, $\Theta^\parallel$ and $\Theta^\perp$, are expressed as
$\Theta^\parallel = -\frac{1}{4k_\mathrm{B}} \sum_n \ab[J_n + \frac{1}{3} K_n - \frac{1}{3} \ab(\Gamma_n + 2\Gamma_n')]$
and
$\Theta^\perp = -\frac{1}{4k_\mathrm{B}} \sum_n \ab[J_n + \frac{1}{3} K_n + \frac{2}{3} \ab(\Gamma_n + 2\Gamma_n')]$, respectively, where the summation runs over all bonds connected to a given site, and their average is given by $\Theta^\mathrm{ave} = \frac{1}{3} \ab(2\Theta^\parallel + \Theta^\perp)$.
The calculated Curie-Weiss temperatures are summarized in Table~\ref{tab:CW_temp}.
We present results for both the 2D model, which includes only in-plane NN, 2NN, and 3NN interactions, and the 3D model, which additionally incorporates three interlayer interactions, i-NN, i-2NN1, and i-2NN2.
In both materials and in both models, the Curie-Weiss temperatures are positive, which can be attributed to dominant ferromagnetic Heisenberg-Kitaev interactions on in-plane NN bonds.
To extract the effect of interlayer interactions, we define $\varDelta \Theta^\mathrm{ave}$ as the difference between the 3D and 2D values of $\Theta^\mathrm{ave}$, and obtain $\varDelta \Theta^\mathrm{ave} \sim -2.71\,\mathrm{K}$ for RuCl$_3$ and $\varDelta \Theta^\mathrm{ave} \sim -3.33\,\mathrm{K}$ for RuBr$_3$.
These results indicate that interlayer interactions enhance antiferromagnetic correlations in both materials, and that the reduction of $\Theta^\mathrm{ave}$ in RuBr$_3$ is greater than that in RuCl$_3$, suggesting that antiferromagnetic correlations due to interlayer interactions are more pronounced in RuBr$_3$.
This trend is consistent with the experimentally inferred stronger antiferromagnetic tendency of RuBr$_3$, although $\Theta^\mathrm{ave}$ does not directly determine the magnetic ordering temperature.
The impact of interlayer interactions on the magnetic ground state will be discussed in Sec.~\ref{sec:aniso}.

Finally, we compare our results for interlayer interactions with those of previous studies.
In Ref.~\cite{Cen2025}, the interlayer NN interactions in RuCl$_3$ are characterized by a single parameter, which determines the strengths of the Heisenberg interaction $J$ and the off-diagonal interactions $\Gamma$ and $\Gamma'$, possibly because of the assumption of pure $j_\mathrm{eff} = \frac 1 2$ states in the perturbative treatment.
Although inversion symmetry imposes the relation $\Gamma = \Gamma'$ for the interlayer NN bond, the relation between $J$ and $\Gamma$ is not fixed by symmetry.
In our calculations, these two parameters are determined independently, since the local basis is obtained from DFT calculations.
It is known that trigonal distortion, which leads to deviations from the pure $j_\mathrm{eff} = \frac 1 2$ states, can significantly affect the effective interactions~\cite{Winter2017review}.
In the present study, contributions from the $e_g$ orbitals are also included, which also leads to quantitative differences from earlier works.

\subsection{Mean-field results for Ru$X_3$}
    \begin{figure}
        \centering
        \includegraphics[width=\linewidth]{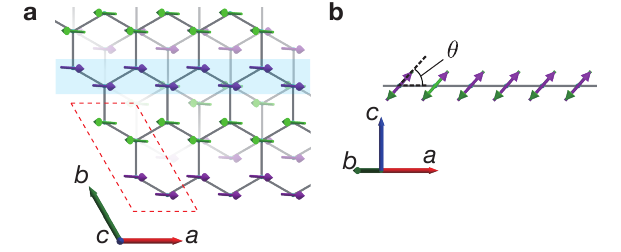}
        \caption{
            \textbf{Mean-field result for the magnetic structure of the pseudospin model of RuCl$_3$}.
            (a) Top view and (b) side view of the mean-field ground state.
            The parallelogram with the red dashed line in panel (a) represents the magnetic unit cell assumed in the calculation.
            To visualize the stacking structure, the magnetic configuration of the upper layer is superimposed in panel (a) with transparency.
            We also show the elevation angle $\theta$ of the magnetic moment measured from the honeycomb plane toward the $c$ axis in panel (b).
            Blue-shaded region in panel (a) is explained in Fig.~\ref{fig:3fZZ_6fZZ}.
        }
        \label{fig:RuCl3_result_mf}
    \end{figure}

    \begin{figure}
        \centering
        \includegraphics[width=\linewidth]{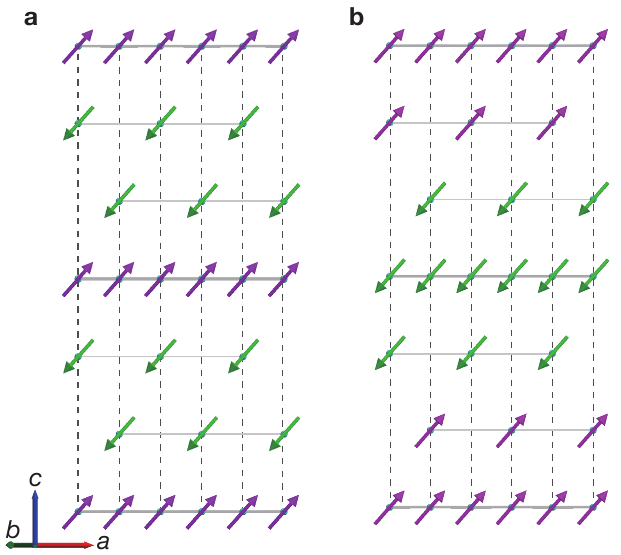}
        \caption{
            \textbf{Calculated stacking patterns of the zigzag magnetic order in RuCl$_3$}.
            Schematic illustrations are shown for (a) the threefold-stacked ZZ state and (b) the sixfold-stacked ZZ state, viewed perpendicular to the $a$--$c$ plane.
            The magnetic structures correspond to the blue-shaded region in Fig.~\ref{fig:RuCl3_result_mf}(a).
            The dashed vertical lines serve as guides to the eye indicating the stacking sequence of the magnetic moments along the $c$ axis.
        }
        \label{fig:3fZZ_6fZZ}
    \end{figure}

In this section, we examine what kind of magnetic order is stabilized by the effective pseudospin models derived in the previous section.
To this end, we analyze the models for RuCl$_3$ and RuBr$_3$ within the mean-field approximation and compare the resulting ordered states with experimental observations.
This comparison serves as a direct test of whether the coupling constants obtained from the first-principles-based construction capture the experimentally relevant magnetic tendencies and the effect of halogen substitution.
In the mean-field calculation, we consider the model including the in-plane NN, 2NN, and 3NN interactions together with the dominant interlayer couplings, i-NN, i-2NN1, and i-2NN2.
For simplicity, we assume a four-sublattice magnetic structure in each honeycomb layer, as indicated by the red dashed parallelogram in Fig.~\ref{fig:RuCl3_result_mf}(a), and compare different stacking patterns along the $c$ direction.
This sublattice choice is sufficient to discuss the stability of FM and ZZ orders, but larger unit-cell calculations are required to examine the possibility of other magnetic orders, such as incommensurate orders; such an analysis is beyond the scope of this study.

Figure~\ref{fig:RuCl3_result_mf} shows the mean-field magnetic structure for RuCl$_3$.
We find that the mean-field ground state of RuCl$_3$ is the three-fold stacked ZZ (3fZZ) order, which is composed of the in-plane zigzag order stacked along the $c$ direction with a three-layer periodicity as shown in Fig.~\ref{fig:3fZZ_6fZZ}(a).
Note that this order pattern is also observed in RuCl$_3$ at low temperatures without magnetic fields.
We also find that another stacking pattern of the in-plane zigzag order, namely, the six-fold stacked ZZ (6fZZ) order [See Fig.~\ref{fig:3fZZ_6fZZ}(b)], is a metastable state with an energy difference of $0.016\,\mathrm{meV}$ per site from the 3fZZ state.
We expect that the 6fZZ order may be stabilized by applying magnetic fields, which has been suggested in Ref.~\cite{Balz2021}.
Although the 3fZZ order is consistent with the experimentally observed magnetic structure of RuCl$_3$, this result differs from the expectation based solely on the classical phase diagram of the NN $JK\Gamma$ model, where the NN exchange constants obtained for RuCl$_3$ are located in the FM region.
This difference indicates that interactions beyond the NN bonds play an essential role in stabilizing the ZZ order in our first-principles-based model.
In particular, the sizable antiferromagnetic 3NN Heisenberg interaction $J_3$, shown in Fig.~\ref{fig:coupling_constant}(d), favors the ZZ order.
We also find an FM solution as a metastable state, with an energy difference of $0.517\,\mathrm{meV}$ per site from the 3fZZ state.

Let us further discuss the obtained ZZ ordering state shown in Fig.~\ref{fig:RuCl3_result_mf}.
For the obtained 3fZZ structure, we define the elevation angle $\theta$ as the angle of the local magnetic moment measured from the honeycomb plane toward the $c$ axis, and obtain $\theta \sim 48^\circ$.
This value is larger than the experimentally observed angle of $\theta \sim 35^\circ$~\cite{Cao2016}, which may be attributed to the overestimation of the ferromagnetic Kitaev interaction $K_1$ in our calculations, as shown in Fig.~\ref{fig:coupling_constant}(b), which tends to cant the ordered magnetic moment out of the honeycomb plane~\cite{Chaloupka2016,Kaib2022}.
With respect to the azimuthal angle of the magnetic moment, our mean-field solution yields that the magnetic moment is almost aligned on the $ac$ plane, which is consistent with previous experimental and theoretical studies~\cite{Cao2016,Chaloupka2016,Suzuki2021,Kaib2022,Park2024}.

    \begin{figure}
        \centering
        \includegraphics[width=\linewidth]{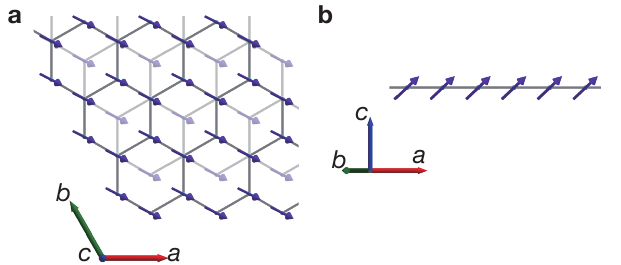}
        \caption{
            \textbf{Magnetic structure for the pseudospin model of RuBr$_3$ within the mean-field theory}.
            Mean-field ground state viewed along (a) the $c$ axis and (b) a direction perpendicular to the $a$ axis.
        }
        \label{fig:RuBr3_result_mf}
    \end{figure}

We next discuss the mean-field results for the magnetic structure of RuBr$_3$.
Figure~\ref{fig:RuBr3_result_mf} shows the mean-field ground state for RuBr$_3$.
In contrast to RuCl$_3$, where the mean-field ground state is the 3fZZ order, we obtain an FM order as the ground state.
Note that we also find a 3fZZ state as a metastable state, which is higher in energy by $0.277\,\mathrm{meV}$ per site.
This result indicates that the FM tendency remains dominant in RuBr$_3$ even after the dominant interlayer interactions are included, even though the Curie-Weiss temperature is reduced in comparison with RuCl$_3$, as discussed in the previous section.
The stabilization of the FM order is likely related to the reduced tendency toward the ZZ phase compared with RuCl$_3$, especially the suppression of the antiferromagnetic 3NN Heisenberg interaction $J_3$ and the modified balance of the NN $JK\Gamma$ interactions, as shown in Fig.~\ref{fig:coupling_constant}(b, d).
On the other hand, previous exact-diagonalization studies of the NN $JK\Gamma$ model with the 3NN Heisenberg interaction indicate that ZZ spin correlations are dominant rather than FM ones for the exchange parameters of RuBr$_3$~\cite{Gretarsson2026}, and experiments establish that the ZZ order is the ground state in RuBr$_3$~\cite{Imai2022, Pearce2024}.
These studies suggest that quantum fluctuations play an important role in stabilizing the ZZ order in RuBr$_3$, or that the present model does not fully capture the interaction balance relevant to the ZZ state.
The latter possibility will be examined in the next section by introducing orbital-dependent deviations from the spherical approximation in the local Coulomb interaction.

\section{Discussion}
\label{sec:aniso}
\begin{figure*}[t]
    \centering
    \includegraphics[width=\linewidth]{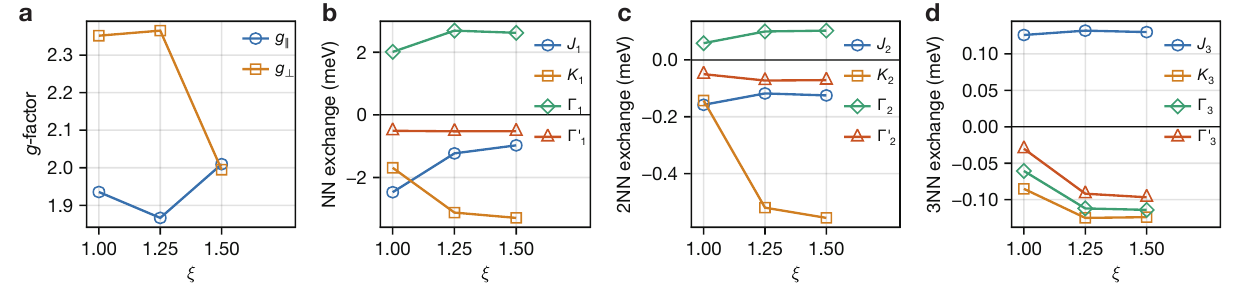}
    \caption{
        \textbf{Calculated parameters of the localized effective model for RuBr$_3$ as a function of the deviation parameter $\xi$ from the spherical approximation for the local Coulomb interaction}.
        (a) $g$-factors and (b-d) coupling constants for (b) NN, (c) 2NN, and (d) 3NN bonds, respectively.
        The coupling constants shown in panels (b-d) are given for the Z-bond indicated in the insets of Fig.~\ref{fig:hopping}(a)--(c).
    }
    \label{fig:coupling_constant_aniso}
\end{figure*}

\begin{figure*}[t]
    \centering
    \includegraphics[width=0.75\linewidth]{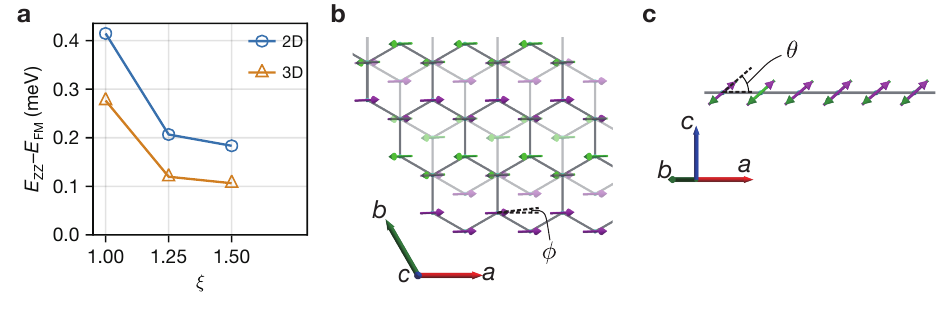}
    \caption{
        \textbf{Effect of the deviation from the spherical approximation for the local Coulomb interaction on the magnetic ground state of RuBr$_3$}
        (a) Energy difference per site between the ZZ mean-field solution and the FM solution for RuBr$_3$ as a function of the deviation parameter $\xi$ from the spherical approximation for the local Coulomb interaction.
        Blue circles and yellow triangles indicate results obtained using only in-plane interactions and including interlayer interactions, respectively.
        (b) Top view and (c) side view of the metastable ZZ ordered state for RuBr$_3$ at $\xi = 1.50$.
        The definitions of the azimuthal angle $\phi$ and the elevation angle $\theta$ are also shown.
    }
    \label{fig:mf_result_br_anisodep}
\end{figure*}

As discussed in the previous section, we found that the mean-field ground state of RuCl$_3$ is the 3fZZ order, which is consistent with experiments, while that of RuBr$_3$ is an FM order, in contrast to the experimentally observed ZZ order.
In the following, we use the mean-field energy difference between the ZZ and FM solutions as a measure of their competition and examine how it is affected by orbital-dependent Coulomb interactions.

In the present study, we constructed a model including all five $d$ orbitals.
For comparison, we also consider a model that includes only the $t_{2g}$ orbitals for the 12 ($=3\times 2\times 2$) bands in the energy window $-1.0 \lesssim E \lesssim 0.3$ in Fig.~\ref{fig:pdos_band}(c), which are commonly considered as the low-energy degrees of freedom in Ru$X_3$.
Performing mean-field calculations for the low-energy localized electron model based on the $t_{2g}$ orbital model, we find that, in contrast to the five-orbital model, its ground state is the ZZ order.
As shown in Fig.~\ref{fig:pdos_band}, RuBr$_3$ has a smaller energy separation between the $t_{2g}$ and $e_g$ orbitals than RuCl$_3$, suggesting that a five-orbital model should be appropriate for RuBr$_3$.
At the same time, the contrasting results of the five-orbital and $t_{2g}$ orbital models indicate that the FM-ZZ competition in RuBr$_3$ is highly sensitive to the contribution of the $e_g$ degrees of freedom.
One possible origin of this sensitivity is that the local Coulomb interaction is treated within the spherical approximation introduced in Sec.~\ref{sec:spherical-approx}.
Since the $t_{2g}$--$e_g$ crystal-field splitting obtained from the first-principles band structure appears to be reasonably estimated, as suggested by its agreement with available spectroscopic energy scales, we focus instead on the local Coulomb interaction, whose orbital dependence is simplified within the spherical approximation.
If this approximation underestimates the energy cost of intermediate-state configurations involving both $t_{2g}$ and $e_g$ electrons, the five-orbital model may overestimate the $e_g$ orbital contribution to the effective exchange interactions.
To examine this effect, we introduce a phenomenological parameter $\xi$ that enhances the direct Coulomb interaction between the $e_g$ and $t_{2g}$ orbitals.
Specifically, we replace the onsite Coulomb matrix elements $U_{\gm\gm'\gm\gm'}$ [see Eq.~\eqref{eq:spherical_int} for definition] by $\xi U_{\gm\gm'\gm\gm'}$ for $\gamma \in e_g$ and $\gamma' \in t_{2g}$, or vice versa, while keeping the other Coulomb matrix elements unchanged.
The original RuBr$_3$ model corresponds to $\xi=1$, and $\xi$ is regarded as a phenomenological parameter for probing the sensitivity to orbital-dependent deviations from the spherical approximation in the local Coulomb interaction.

Figure~\ref{fig:coupling_constant_aniso}(a) shows $\xi$ dependence of the $g$-factors.
In the original model ($\xi = 1$), the calculated $g$-factors exhibit easy-axis anisotropy as shown in Fig.~\ref{fig:coupling_constant}(a), opposite to experiments, whereas at $\xi = 1.50$ they approach easy-plane anisotropy and become nearly isotropic.
This indicates that the magnetic anisotropy is highly sensitive to orbital-dependent deviations from the spherical approximation in the local Coulomb interaction.

We next discuss the impact of the $\xi$-induced deviation from the spherical approximation on the coupling constants.
Figures~\ref{fig:coupling_constant_aniso}(b-d) show the $\xi$ dependence of the in-plane coupling constants of the localized effective model, as in Fig.~\ref{fig:coupling_constant}.
We find that enhancing the $e_g$--$t_{2g}$ interorbital Coulomb interaction through $\xi$ causes a change in the relative magnitudes of the dominant NN coupling constants, namely the Heisenberg interaction $J_1$ and the Kitaev interaction $K_1$.
Moreover, according to the phase diagrams of the $JK\Gamma$ model~\cite{Rau2014}, the system is located in the in-plane FM phase at $\xi=1$, but moves toward the ZZ phase as $\xi$ increases.

We then perform a mean-field analysis for the localized model obtained with the $\xi$-induced deviation from the spherical approximation for RuBr$_3$.
We find that the FM solution remains the ground state for $\xi$ up to 1.50, while the ZZ solution becomes metastable as in the original model with $\xi=1$.
Figure~\ref{fig:mf_result_br_anisodep}(a) shows the energy difference per site between the metastable ZZ state and the FM ground state as a function of $\xi$.
The blue circles and yellow triangles correspond to the 2D and the 3D models defined in Sec.~\ref{sec:spin_model}, respectively.
In both the 2D and the 3D models, the energy difference between the ZZ and FM states decreases as $\xi$ increases, indicating that the $\xi$-induced deviation enhances the tendency toward the ZZ state.
We also find that the energy difference in the 3D model is smaller than that in the 2D model for all $\xi$.
This is consistent with the discussion in Sec.~\ref{sec:spin_model} that interlayer couplings enhance antiferromagnetic correlations.

Here, we analyze the effect of the $\xi$-induced deviation on the magnetic structure of the metastable ZZ state in more detail.
Figures~\ref{fig:mf_result_br_anisodep}(b, c) show the magnetic structure of the metastable mean-field state at $\xi=1.50$, viewed along the $c$ axis and along a direction perpendicular to the $a$ axis, respectively.
The mean-field analysis shows that the elevation and azimuthal angles of the magnetic moment are $\theta \sim 41^\circ$ and $\phi \sim 3^\circ$ for this metastable ZZ state, respectively, while the original model based on the spherical approximation ($\xi = 1$) gives $\theta \sim 59^\circ$ and $\phi \sim 34^\circ$ for the corresponding ZZ state.
This result suggests that the $\xi$-induced deviation can significantly affect the azimuthal direction of the magnetic moment and drive the magnetic moment toward the $ac$ plane.
Indeed, neutron-scattering experiments report that the ZZ magnetic order in RuBr$_3$ has an elevation angle and azimuthal angle of approximately $\theta \sim 60^\circ$ and $\phi \sim 0^\circ$, respectively~\cite{Imai2022}.
The mean-field result for the metastable ZZ state in the model with the $\xi$-induced deviation from the spherical approximation is consistent with this experimental result with respect to the azimuthal angle $\phi$.
Note that the experimentally observed elevation angle $\theta$ for the ZZ state is closer to that of the metastable ZZ state in the model with $\xi=1$ rather than that with $\xi=1.5$, which might be attributed to factors that cannot be captured by the parameter $\xi$.

Finally, we comment on the effect of the same $\xi$-induced deviation for RuCl$_3$.
As $\xi$ increases, in contrast to RuBr$_3$, the mean-field ground state changes from 3fZZ to FM.
This material-dependent response may be related to the situation suggested by RIXS measurements in RuCl$_3$, where strong competition exists between FM and ZZ orderings~\cite{Suzuki2021}.
See supplementary information for detailed results.

Thus, although the ZZ state does not become the mean-field ground state within the examined range of $\xi$, the systematic reduction of $E_\mathrm{ZZ}-E_\mathrm{FM}$ demonstrates that orbital-dependent Coulomb interactions shift the magnetic competition toward the ZZ state and strongly affect the magnetic anisotropy.
Another possible origin of the remaining discrepancy is quantum fluctuations beyond the present mean-field treatment; indeed, previous theoretical studies have suggested that quantum fluctuations can stabilize the ZZ order in RuBr$_3$ despite a smaller third-neighbor Heisenberg interaction than in RuCl$_3$~\cite{Kaib2022}, consistent with the present results shown in Fig.~\ref{fig:coupling_constant}(d).
Including quantum fluctuations in the model with the $\xi$-induced deviation from the spherical approximation is an important direction for future work to further clarify the origin of the ZZ order in RuBr$_3$.

In summary, we have investigated the effect of halogen substitution on the magnetic properties of the Kitaev candidate materials Ru$X_3$ ($X=$ Cl, Br) through a first-principles-based analysis.
We constructed effective pseudospin models using perturbation theory from the strong-coupling limit of multiorbital Hubbard models whose parameters were determined from first-principles calculations, and examined differences in the magnetic structures of RuCl$_3$ and RuBr$_3$.
The first-principles calculations show that the Wannier functions for RuBr$_3$ are more spatially extended than those for RuCl$_3$, reflecting the smaller charge-transfer gap and enhanced Ru $4d$--ligand $p$ hybridization in RuBr$_3$.
This leads to larger transfer integrals and smaller onsite Coulomb interactions in RuBr$_3$.
In addition, the $t_{2g}$--$e_g$ orbital splitting is smaller in RuBr$_3$ than in RuCl$_3$, suggesting an enhanced contribution of $e_g$ orbitals to the magnetic properties of RuBr$_3$.

The obtained exchange interactions indicate that the antiferromagnetic third-neighbor Heisenberg interaction is suppressed in RuBr$_3$, whereas interlayer exchange interactions are enhanced, reflecting the larger spatial extent of the Wannier functions.
These enhanced interlayer interactions strengthen three-dimensional correlations, consistent with the stronger antiferromagnetic tendency experimentally inferred for RuBr$_3$.
By performing a mean-field analysis for the effective pseudospin models, we find that RuCl$_3$ realizes the zigzag order stacked with a three-layer periodicity along the $c$ axis, with the magnetic moment almost aligned on the $ac$ plane, consistent with experiments.
On the other hand, the mean-field ground state of RuBr$_3$ is ferromagnetic, in contrast to the experimentally observed zigzag order, while a zigzag solution is obtained as a metastable state lying only $0.277\,\mathrm{meV}$ per site above the ground state.
To clarify the origin of this discrepancy, we analyzed a model incorporating orbital-dependent deviations from the spherical approximation for the local Coulomb interaction.
This phenomenological analysis shows that the magnetic anisotropy and the energy balance between ferromagnetic and zigzag states are highly sensitive to such deviations, and that the magnetic moment in the metastable zigzag state is driven toward the experimentally observed azimuthal direction.

These results demonstrate that halogen substitution controls magnetic competition in Ru$X_3$ through both three-dimensional exchange interactions and effects of orbital-dependent Coulomb interactions.
In the present study, we have adopted the spherical approximation for the local Coulomb interaction, which is commonly used in first-principles-based analyses of correlated materials, and have phenomenologically introduced deviations from this approximation to examine their effects on the magnetic properties of RuBr$_3$.
Beyond this phenomenological analysis, it would be desirable to evaluate the four-center Coulomb integrals based on Wannier functions with high precision, although this approach is extremely expensive in the current computational framework.
To overcome this difficulty, imposing symmetry constraints consistent with the site symmetry of the material may be a promising approach to reduce the computational cost~\cite{Iimura2021}.
In addition, we have analyzed the magnetic properties of Ru$X_3$ within the mean-field approximation, which neglects quantum fluctuations.
Including quantum fluctuations in the model is therefore an important direction for future work to further clarify the origin of the zigzag order in RuBr$_3$ and to establish a microscopic understanding of halogen-substituted Kitaev materials.

\section{Methods}
\subsection{DFT calculation and Wannierization}
    DFT calculations were performed using \texttt{Quantum ESPRESSO}~\cite{Giannozzi2017}.
    For both RuCl$_3$ and RuBr$_3$, we adopted the $R\bar 3$ crystal structure reported in Refs.~\cite{Park2024, Imai2022}.
    For RuBr$_3$, the structure determined at 3\,K was used.
    All calculations were performed for the rhombohedral structure.
    Norm-conserving pseudopotentials within the revised Perdew-Burke-Ernzerhof exchange-correlation functional~\cite{Hamann2013} were taken from the \texttt{Pseudo Dojo} library~\cite{vanSetten2018}.
    The plane-wave energy cutoff was set to 100\,Ry, and a $\bm k$-point mesh of $6\times 6\times 6$ was employed for self-consistent-field calculation.
    Wannier functions were constructed using \texttt{Wannier90}~\cite{Pizzi2020} without performing maximal localization.
    As initial projections, we selected the Ru-site $d$-orbitals.

    For the $t_{2g}$ orbital model of RuBr$_3$ introduced in Sec.~\ref{sec:aniso}, we restricted the Wannierization to the low-energy bands within the energy window from $-1.0$ to $0.3\,\mathrm{eV}$.
    The Ru $d_{yz}$, $d_{zx}$, and $d_{xy}$ orbitals were used as the initial projections for constructing the Wannier functions.

\subsection{Spherical approximation for Coulomb interaction}
\label{sec:spherical-approx}
    The Coulomb interaction was evaluated using \texttt{RESPACK}~\cite{Nakamura2021} within the cRPA.
    Since the current version of \texttt{RESPACK} does not support fully relativistic pseudopotentials, the Wannier functions used for evaluating the interaction parameters were constructed with scalar-relativistic pseudopotentials under the same computational setup as in the previous subsection.
    The interface \texttt{wan2respack}~\cite{Kurita2023} was employed to run \texttt{RESPACK} based on the \texttt{Wannier90} output.

    The four-center integrals $U_{\gm_1 \gm_2 \gm_3 \gm_4}$ in the multiorbital Hubbard model were evaluated within the spherical approximation using the direct and exchange integrals calculated by \texttt{RESPACK}:
    \begin{align} \label{eq:direct_int}
        &U_{\gm\gm'} =
        \iint \d{\bm r} \d{\bm r'} w_\gm^*(\bm r) w_{\gm'}^*(\bm r') W(\bm r, \bm r') w_\gm(\bm r) w_{\gm'}(\bm r')
        \\ \label{eq:exch_int}
        &J_{\gm\gm'} =
        \iint \d{\bm r} \d{\bm r'} w_\gm^*(\bm r) w_{\gm'}^*(\bm r') W(\bm r, \bm r') w_{\gm'}(\bm r) w_\gm(\bm r')
    \end{align}
    where $w_\gm(\bm r)$ is the Wannier function for orbital $\gm$, and $W(\bm r, \bm r')$ is the screened Coulomb interaction obtained within cRPA.

    According to ligand-field theory, the spherically symmetric interaction between electrons with the same azimuthal quantum number can be expressed in terms of the Slater-Condon parameters $F^p$ as
    \begin{align} \label{eq:spherical_int}
        U_{\gm_1\gm_2\gm_3\gm_4} = \sum_p F^p c^p(\gm_1, \gm_3) c^p(\gm_2, \gm_4),
    \end{align}
    where $c^p$ are the Gaunt coefficients, given by the angular integrals of products of spherical harmonics~\cite{SuganoBook}.
    The summation over $p$ is restricted by the Gaunt coefficients; for a $d$-electron system, only $p=0, 2, 4$ are finite.
    Thus, in the $d$-electron case, the interaction tensor $U_{\gm_1\gm_2\gm_3\gm_4}$ is fully determined by three independent parameters, $F^0, F^2,$ and $F^4$.

    In the following, we focus on the $d$-electron case.
    From the direct and exchange integrals calculated in Eqs.~\eqref{eq:direct_int} and \eqref{eq:exch_int}, we define the average intra-orbital direct integral, the average exchange integral within the $t_{2g}$ orbitals, and the exchange integral between the $e_g$ orbitals as
    \begin{align}
        &U = \frac 1 5 \sum_\gm U_{\gm\gm},
        \\
        &J_{t_{2g}} = \frac 1 3 \ab(J_{yz,zx} + J_{zx,xy} + J_{xy,yz}),
        \\
        &J_{e_g} = J_{z^2,x^2-y^2}.
    \end{align}
    These quantities can be expressed in terms of the Slater-Condon parameters as
    \begin{align}
        &U = F^0 + \frac{4}{49} F^2 + \frac{4}{49} F^4,
        \\
        &J_{t_{2g}} = \frac{3}{49} F^2 + \frac{20}{441}F^4,
        \\
        &J_{e_g} = \frac{4}{49}F^2 + \frac{5}{147}F^4,
    \end{align}
    from which we determine $F^0, F^2,$ and $F^4$.
    Using these parameters, the spherically symmetric interaction in Eq.~\eqref{eq:spherical_int} is evaluated.

\subsection{Perturbation theory}
    Starting from the multiorbital Hubbard model constructed in the previous subsections,
    \begin{align}
        &\mathscr H = \mathscr H_t + \mathscr H_\mathrm{loc},
        \\
        &\mathscr H_t =
        \sum_{\la ij \ra} \sum_{\gm\gm'} \sum_{\sg\sg'} t_{ij}^{\gm\sg,\gm'\sg'} c_{i\gm\sg}^\dg c_{j\gm'\sg'} + \Hc,
        \\
        &\mathscr H_\mathrm{loc} =
        \mathscr H_\mathrm{CEF} + \mathscr H_\mathrm{SOC}
        \nonumber
        \\ &\hspace{.5cm}
        + \frac 1 2 \sum_i \sum_{\gm_1\gm_2\gm_3\gm_4} \sum_{\sg\sg'} U_{\gm_1\gm_2\gm_3\gm_4} c_{i\gm_1\sg}^\dg c_{i\gm_2\sg'}^\dg c_{i\gm_4\sg'} c_{i\gm_3\sg},
    \end{align}
    we derive a localized effective model by performing a perturbative expansion in the strong coupling limit ($\mathscr H_\mathrm{loc} \gg \mathscr H_t$)~\cite{Winter2016,Winter2017review,Kaib2022}.
    Here, $\mathscr H_\mathrm{CEF}$ and $\mathscr H_\mathrm{SOC}$ denote the local crystalline electric field and the spin-orbit coupling obtained from \texttt{Wannier90}.

    \begin{figure}
        \centering
        \includegraphics[width=\linewidth]{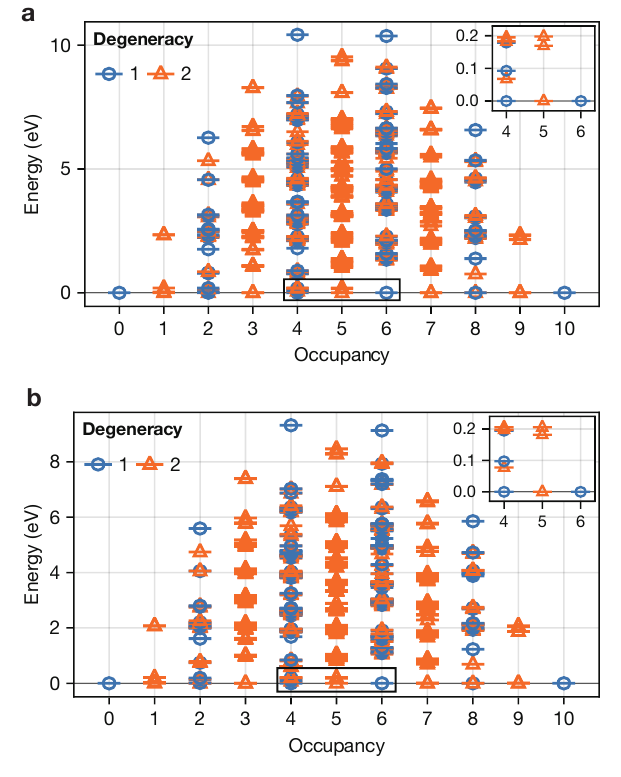}
        \caption{
            \textbf{Calculated energy levels of the local Hamiltonian for RuCl$_3$ and RuBr$_3$}.
            (a) Energy levels of the local Hamiltonian for RuCl$_3$.
            The horizontal axis represents the number of electrons per site, and the vertical axis shows the energy measured from the lowest level for each filling.
            The color and shape of the markers indicate the degeneracy of each level.
            The inset shows an enlarged plot for the low-energy region enclosed by the black square.
            (b) Corresponding energy levels for RuBr$_3$.
        }
        \label{fig:ene_level}
    \end{figure}

    We first diagonalize the local Hamiltonian $\mathscr H_\mathrm{loc}$ to obtain its eigenvalues and eigenstates,
    \begin{align}
        \mathscr H_i \ket|\al_i> = E_{\al_i} \ket|\al_i>,
    \end{align}
    where $\mathscr H_i$ denotes the onsite part of $\mathscr H_\mathrm{loc}$ at site $i$.
    The resulting energy levels for each electron number are shown in Fig.~\ref{fig:ene_level}.
    The horizontal axis indicates the local electron number, and the vertical axis shows the energy measured from the lowest level for each filling.
    The color and marker type represent the degeneracy of each level.
    The inset shows enlarged plots of the low-energy region.
    All levels with odd electron numbers are doubly degenerate due to Kramers degeneracy associated with SOC.
    Based on this energy-level structure, we define the model Hilbert space for constructing the effective model.
    In this work, we choose the lowest doublet of the $d^5$-configuration as the model Hilbert space.
    In the absence of trigonal distortion, this doublet corresponds to the $j_\mathrm{eff}=\frac 1 2$ state of the $t_{2g}^5$-configuration.
    However, since we include trigonal distortion and all five $d$-orbitals, the resulting doublet is not a pure $j_\mathrm{eff}=\frac 1 2$ state as mentioned in main text.

    The matrix elements of the second-order effective Hamiltonian are given by
    \begin{align}
        &\braket*<\!\braket*<\bm \al|\mathscr H_\mathrm{eff}^{(2)}|\bm \beta>\!>
        \nonumber
        \\
        &= \frac 1 2 \sum_{\bm \al'\neq \bm \al, \bm \beta} \braket*<\!\braket*<\bm \al|\mathscr H_t|\bm \al'>\!> \ab[
            \frac{1}{E_{\bm \al} - E_{\bm \al'}}
            + \frac{1}{E_{\bm \beta} - E_{\bm \al'}}
        ] \braket*<\!\braket*<\bm \al'|\mathscr H_t|\bm \beta>\!>,
    \end{align}
    where $\ket*|\bm \al>\!\ra = \prod_i \ket*|\al_i>$ denotes a many-body state.
    The calculated effective Hamiltonian is then expanded in terms of pseudospin operators defined within the two-state model Hilbert space,
    \begin{align}
    \left(\tilde S_i^x,\tilde S_i^y,\tilde S_i^z\right)=\frac12\left(\sg_i^x,\sg_i^y,\sg_i^z\right),
    \end{align}
    leading to the localized effective model
    \begin{align} \label{eq:H_eff}
        \mathscr H_\mathrm{eff} = \sum_{\la ij \ra}\sum_{\mu\nu} J_{ij}^{\mu\nu} \tilde S_i^\mu \tilde S_j^\nu,
    \end{align}
    as described in Refs.~\cite{Iwazaki2021, Iwazaki2023}.
    Here, $\sg_i^x, \sg_i^y, \sg_i^z$ are Pauli matrices, which represent the matrix elements in the subspace spanned by the twofold-degenerate local many-body states $\ket|\al_i>$ at site $i$.

    In this study, we constructed the pseudospin model by means of perturbation theory with respect to the intersite hopping of the multiorbital Hubbard model.
    As the initial and final states in the perturbation theory, we numerically diagonalized the local Hamiltonian consisting of the onsite Coulomb interaction, CEF, and SOC, and selected its eigenstates.
    In the present procedure, however, the pseudospin quantization axis becomes ambiguous.
    In this paragraph, we describe how we estimated the parameters $J, K, \Gamma,$ and $\Gamma'$ of the localized effective model in such a situation.
    Since RuCl$_3$ and RuBr$_3$ belong to the space group $R\bar 3$, we choose the local pseudospin coordinate system such that the crystallographic $c$ axis corresponds to the pseudospin $[111]$ direction.
    The remaining rotational freedom about the $[111]$ axis is fixed so that the coupling matrix for the in-plane NN Z-bond takes the standard $JK\Gamma\Gamma'$ form of Eq.~\eqref{eq:Jmel}.
    In practice, the rotation angle is determined by minimizing $|J^{xx}-J^{yy}|$.

    We next consider the magnetic-field response of the localized effective model.
    The magnetic moment can be written in terms of the orbital moment $\bm L_i$ and spin $\bm S_i$ as
    \begin{align}
        \bm M_i = -\bm L_i - 2\bm S_i.
    \end{align}
    Projecting the magnetic moment onto the model Hilbert space, we obtain
    \begin{align} \label{eq:rel_mag2pspin}
        \mathscr P_i M_i^\mu \mathscr P_i = \sum_\nu g_i^{\mu\nu} \tilde S_i^\nu,
    \end{align}
    where $\mathscr P_i = \sum_{\al_i} \ketbra*|\al_i><\al_i|$ is the projection operator onto the model Hilbert space.
    The expansion coefficients define the $g$-factors,
    \begin{align}
         g_i^{\mu\nu} = 2 \sum_{\al_i} \braket*<\al_i|M_i^\mu \tilde S_i^\nu|\al_i>.
    \end{align}
    In this way, the response to an external magnetic field in the localized effective model is described through the correspondence between physical operators and their representation in the model Hilbert space.
    For convenience, the $g$-tensor presented in the main text is represented in an orthogonal coordinate system containing the crystallographic $a$ and $c$ axes shown in Fig.~\ref{fig:crystal_structure}.

\subsection{Mean-field theory}
    In the following, we analyze the localized effective model in Eq.~\eqref{eq:H_eff}, constructed from first-principles calculations, within the mean-field approximation.
    Applying the mean-field decoupling to Eq.~\eqref{eq:H_eff}, we obtain
    \begin{align} \label{eq:H_MF}
        \mathscr H_\mathrm{MF}
        =& -\sum_i \sum_\mu
        \ab[
            -\sum_{j \neq i} \sum_\nu J_{ij}^{\mu\nu} \la\tilde S_j^\nu\ra
        ] \tilde S_i^\mu
        \nonumber
        \\
        &- \sum_{\la ij \ra}\sum_{\mu\nu} J_{ij}^{\mu\nu} \la\tilde S_i^\mu\ra \la\tilde S_j^\nu\ra,
    \end{align}
    where $\la\cdots\ra$ denotes the expectation value taken with respect to the mean-field ground state.
    The expectation value of the magnetic moment $\bm M_i$ is obtained by transforming the pseudospin expectation values using Eq.~\eqref{eq:rel_mag2pspin}.

\section*{Acknowledgement}

Parts of figures are drawn by using \texttt{Makie.jl}~\cite{Makie}.
R.I. would like to thank Y.~Nomura and T.~Miki for advice on evaluating the Coulomb interaction.
The authors acknowledge Y.~Motome, Y.~Imai, H.~Suzuki, A.~Ono, and S.M.~Winter for fruitful discussions.
This work was supported by KAKENHI Grants
No.~JP23H04865, No.~JP23H04869, No.~JP24K00563, No.~JP25K23345, No.~JP26H00624, No.~JP26K00652.

\section*{Author contributions}
    R.I. and J.N. conceived the project and wrote the manuscript.
    R.I. performed all of the calculations.
    The numerical code for mean-field calculation was implemented by S.K.
    All authors discussed the results and commented on the manuscript.

\section*{Competing interests}
    The authors declare no competing interests.

\bibliography{main.bbl}

\clearpage

\appendix

\makeatletter
\renewcommand{\thepage}{S\arabic{page}}
\renewcommand{\theequation}{S\arabic{equation}}
\renewcommand{\thefigure}{S\arabic{figure}}
\renewcommand{\thetable}{S\arabic{table}}
\makeatother

\setcounter{page}{1}
\setcounter{equation}{0}
\setcounter{table}{0}
\setcounter{figure}{0}

\noindent
{\bf SUPPLEMENTARY INFORMATION FOR \\
``Halogen control of magnetic competition in Kitaev candidate Ru$X_3$ ($X =$ Cl, Br)''}
\\[2mm]
R. Iwazaki, S. Koyama, T. Koretsune, S. Hoshino, and J. Nasu
\\[2mm]
(Dated: \today)

\section*{SI 1: Effect of deviation from spherical Coulomb interaction on $\text{RuCl}_3$}
    \begin{figure*}[t]
        \centering
        \includegraphics[width=\linewidth]{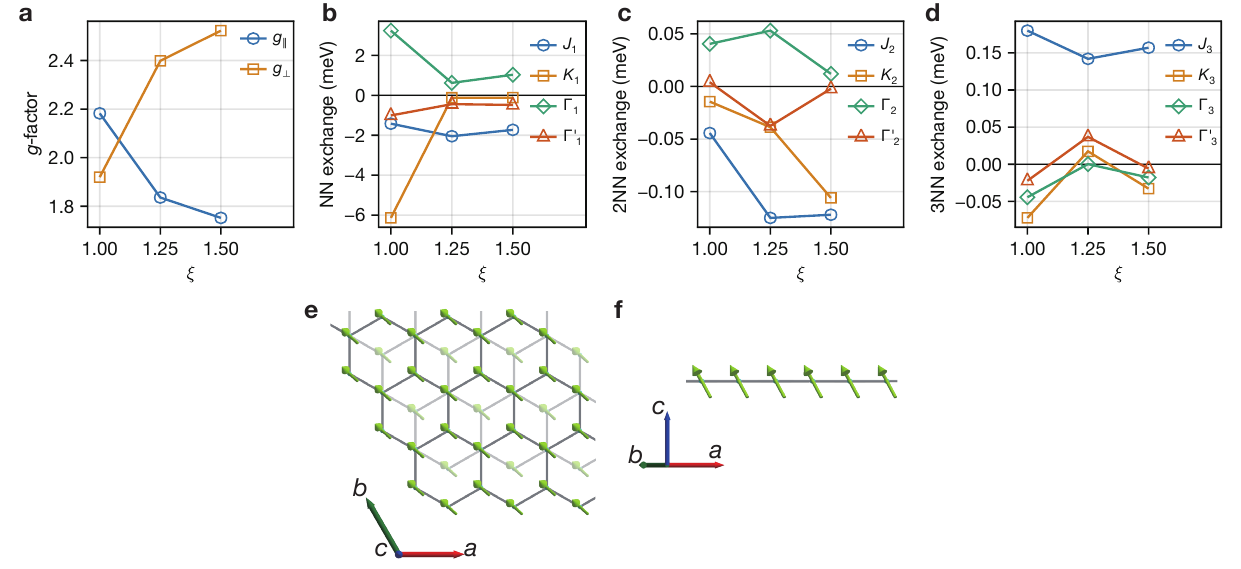}
        \caption{
            \textbf{Calculated parameters of the localized effective model and the corresponding mean-field ground state for RuCl$_3$}
            (a) $g$-factors and (b-d) coupling constants for (b) NN, (c) 2NN, and (d) 3NN bonds, respectively.
            The coupling constants shown in panels (b-d) are given for the Z-bond indicated in the insets of Fig.~\ref{fig:hopping}(a)--(c) in the main text.
            (e, f) Schematic illustrations of the mean-field ground state at $\xi = 1.50$, viewed along (e) the $c$ axis, and (f) perpendicular to the $a\text{--}c$ plane.
        }
        \label{fig:coupling_constant_aniso_cl}
    \end{figure*}

    In this section, we discuss the effect of deviations from the spherical approximation for the local Coulomb interaction in RuCl$_3$.
    We perform the same analysis as that presented for RuBr$_3$ in the main text.

    Figs.~\ref{fig:coupling_constant_aniso_cl}(a)--(d) show the dependence of the parameters of the pseudospin model on the deviation parameter $\xi$ for RuCl$_3$.
    For the $g$-factors shown in Fig.~\ref{fig:coupling_constant_aniso_cl}(a), we find that increasing $\xi$ enhances the easy-axis anisotropy ($g_\perp > g_\parallel$), in contrast to RuBr$_3$, where the magnetic anisotropy becomes nearly isotropic as $\xi$ increases.
    Figures~\ref{fig:coupling_constant_aniso_cl}(b)--(d) summarize the $\xi$ dependence of the coupling constants for the in-plane (b) nearest-neighbor (NN), (c) second-nearest-neighbor, and (d) third-nearest-neighbor bonds.
    For RuCl$_3$, the relative magnitudes of the NN Kitaev interaction $K_1$ and Heisenberg interaction $J_1$ are reversed as $\xi$ increases, indicating a substantial modification of the balance among the dominant magnetic interactions.

    Figs.~\ref{fig:coupling_constant_aniso_cl}(e) and (f) show the mean-field ground state of RuCl$_3$ at $\xi = 1.50$, viewed along the $c$ axis and perpendicular to the $a$--$c$ plane, respectively.
    While the original model with $\xi = 1.00$ exhibits the threefold-stacked zigzag (3fZZ) state as its ground state, the model with finite deviation from the spherical approximation stabilizes a ferromagnetic (FM) ground state.
    Furthermore, we find that the ZZ state is no longer obtained even as a metastable solution for $\xi = 1.25$ and $1.50$.
    This behavior is likely related to the suppression of the NN Kitaev interaction relative to the Heisenberg interaction and the resulting modification of the competition among the dominant exchange couplings.
    Indeed, the effective NN parameters for $\xi = 1.25$ and $1.50$ are located within the FM region of the $JK\Gamma$ phase diagram~\cite{Rau2014}, consistent with the mean-field results.

    Taken together with the corresponding results for RuBr$_3$ presented in the main text, these findings demonstrate that the magnetic ground state is highly sensitive to orbital-dependent deviations from the spherical approximation of the local Coulomb interaction.
    While increasing $\xi$ stabilizes the ZZ state relative to the FM state in RuBr$_3$, the opposite tendency is found in RuCl$_3$, where the ZZ state is destabilized and eventually disappears as a metastable solution.
    These opposite responses indicate that the balance between FM and ZZ magnetic orders is strongly controlled by the details of the local Coulomb interaction.
    This observation is consistent with the competition between FM and ZZ magnetic structures suggested by resonant inelastic X-ray scattering experiments~\cite{Suzuki2021}.

\vspace{10mm}
\noindent
{\bf \large References}
\\[1mm]
See the list of references in the main text.

\end{document}